\documentstyle[preprint,aps]{revtex}
\tightenlines

\begin{document}
\newcommand{\beq}{\begin{equation}}
\newcommand{\eeq}{\end{equation}}
\newcommand{\beqa}{\begin{eqnarray}}
\newcommand{\eeqa}{\end{eqnarray}}
\newcommand{\fr}{\frac}
\draft
\preprint{INJE-TP-01-07, hep-th/0108092}

\title{ Quintessence with a localized scalar field on the brane}

\author{ Y.S. Myung\footnote{Email-address :
ysmyung@physics.inje.ac.kr}}
\address{Relativity Research Center and School of Computer Aided Science, Inje University, Kimhae 621-749, Korea}

\maketitle

\begin{abstract}
We study  issues of the  quintessence in the brane cosmology.
The initial bulk spacetime consists of  two 5D topological anti de Sitter
black hole joined by the brane (moving domain wall).
Here we do not introduce any conventional radiation and matter.
Instead we include a localized scalar  on the brane as a stress-energy tensor,
and thus  we find  the quintessence  which gives an accelerating
universe. Importantly, we obtain a $\rho^2$-term as well as
a holographic matter term of $\alpha/a^4$ from the masses of the
topological black holes. We discuss a possibility that in the early universe,
$\rho^2$-term makes a large kinetic term which induces a
decelerating universe. This may provide a hint of avoiding from the
perpetually accelerating universe of the present-day quintessence.
If a holographic matter term exists, it will plays the role of
a CFT-radiation
in the early universe.
\end{abstract}
\vfill
Compiled at \today : \number \time.

\newpage
\section{Introduction}
\label{introduction}
Recently an accelerating universe has proposed to be a way
to interpret the astronomical data of supernova\cite{Per,CDS,Gar}.
Combining these data with the measurements  of the cosmic microwave background,
it may be led that the universe has  the critical density which consists of
1/3 of ordinary matter and 2/3 of dark matter with a negative pressure as
$p<-\rho/3 $. At this moment a promising candidate for the dark matter
is either a positive cosmological constant or a slowly evolving  scalar known as
``quintessence". Also the quintessence  can  be considered
as  an alternative method to resolve the cosmological constant problem.
This is possible because instead of the fine-tuning, the quintessence  provides a
 model of slowly decaying cosmological constant.

However, there exists another question for the present-day quintessence. This is to ask
whether the expansion will keep accelerating forever or it will
decelerate  again after some time. This is very similar to the
exit problem in the inflationary cosmology.
Some literature  discussed on this issue within the string
theory\cite{Wit,HKS,FKMP} and  realistic cosmological context\cite{RP,KLa}.
If the expansion keeps accelerating eternally, we have the universe
with an event horizon which gives arise to a serious problem in defining the
conventional S-matrix. Also it is hard to obtain de Sitter space
(its asymptotic space) from  the string theory.

On the other hand authors in ref.\cite{CL} also considered the possibility of quintessence
in the  dilatonic domain wall. After the integration over the dual holographic
field theory, one obtained an effective dilaton gravity with the potential on the brane.
They used
the Hamilton-Jacobi method inspired by the holographic
renormalization group to investigate the intrinsic
Friedmann-Robertson-Walker (FRW) cosmology on the brane.
It was shown that a holographic
quintessence is allowed on the dilatonic brane
because a single Liouville-type potential appears on the brane.
However, using the zero-mode approach leads to the fact that
the dilatonic domain wall including the Randall-Sundrum (RS) model
\cite{RS2} cannot accommodate the quintessence\cite{myung}.
 Although we found Liouville-type potentials, we failed to find any
accelerating universe from the zero mode approach of the dilatonic domain wall.
 This is mainly because the
unstable potential was found.

In this paper we wish to deal with the same issue within the brane
cosmology context. More recently the related issues were discussed
in\cite{HLI,MMA}.
 The brane cosmology   contains  two important
deviations from the standard Friedmann-Robertson-Walker (FRW) cosmology :
The first is that  $\rho^2$-term appears in the  Friedmann
equations\cite{BDL,CGKT,KRA}. The second is that there exist a holographic matter term
of $\alpha/a^4$, which is interpreted as either a dark matter\cite{BDL,KRA,MMA} or
a CFT-matter\cite{VER}.
Finally the acceleration or deceleration  of
evolving universe will be determined by a more complicated fashion
than a standard cosmology.  Hence we expect to find something new to
avoid   the perpetually accelerating universe.
Here we study mainly how the brane cosmology with
 a localized scalar and a holographic matter affects
the quintessence of a standard scalar model.

There are two approaches in the brane cosmology.
One  approach is first to assume the 5D dynamic metric (that is, Binetruy-Deffayet-Longlois
metric\cite{BDL})
 which is
manifestly $Z_2$-symmetric under $\hat w \to -\hat w
$\footnote{ $\hat w$ is the fifth one of Gaussian normal coordinates
 $\{\hat \tau,\hat x_i, \hat w \}$.}.
Then one considers the Israel junction
condition to take into account a localized matter distribution on the brane \cite{ISR} and
solves the bulk Einstein equation to find the behavior of the scale factor on the brane.
 We call this   the BDL approach.
The other approach starts with
a static  bulk configuration which consists of  two
5D anti de Sitter-Schwarzschild (AdSS$_5$) black hole spaces joined by the domain wall.
 In this case,
embedding  the moving domain wall (MDW) into the bulk spacetime
is possible by choosing an appropriate
normal vector $n_M$ and tangent vector $t_M$\cite{KRA}.
The domain wall separating two such bulk spaces is taken to be
located at $r=a(\tau)$, where $a(\tau)$ will determined by solving
solely the Israel junction condition. Then observers on the MDW will
interpret their motion through the static bulk background as
cosmological expansion or contraction. This is called the MDW approach.
Although these two approaches seem to be different,
these give us the same result. Actually Mukhoyama et al.\cite{MSM} performed
a coordinate transformation $\{\hat \tau,\hat x_i,\hat w \}$ $\to$
$\{t(\tau),a(\tau),\chi,\theta,\phi\}$ in order to bring the BDL metric into
the AdSS$_5$-metric.

The organization of our paper is as follows. In Sec. II we briefly
review a simple model for the quintesence. We study issues of
the quintessnce within the brane cosmology
in Sec. III.  Finally we discuss our results in Sec.IV.

\section{Quintessence}
A minimally coupled
scalar field with a potential that decreases as the field
increases is usually introduced for the quintessence\cite{RP}.
  The corresponding action is given
by\footnote{Our convention is $(-+++)$.}
\beq
S_{\rm Q} = {1 \over 2} \int d^4 x \sqrt{-g} \left [
 \fr{1}{\kappa_4^2}R - (\partial \phi)^2 -2 V(\phi)\right]
\label{Qact}
\eeq
with the 4D gravitational   constant $\kappa_4^2=8\pi G_N$.
Hereafter we choose $\kappa_4^2=1$ for simplicity.
This is a canonically normalized scalar action coupled to gravity.
The Einstein equation is
\beq
R_{\mu\nu}- \frac{1}{2}R g_{\mu\nu}= T_{\mu\nu}
\label{Qeq}
\eeq
with
\beq
T_{\mu\nu}= \partial_\mu\phi \partial_\nu\phi
-\frac{1}{2}(\partial \phi)^2g_{\mu\nu} -V(\phi)g_{\mu\nu}.
\label{Qst}
\eeq
Considering the FRW universe of $ds^2_{FRW}=-dt^2
+a^2(t)d\Sigma^2_k$,  the equation of motion for $\phi$ and the
conservation law of $\nabla_\mu T^{\mu0}=0$ lead to the the same
equation as
\beq
\ddot \phi + 3\frac{\dot a}{a} \dot \phi +
V'(\phi)=0.
\label{seqa}
\eeq
 The two FRW equations are given by
\begin{equation}
\frac{\dot a^2}{a^2} = -\fr{k}{a^2}+ \frac{\rho}{3},~~~ \frac{\ddot a}{a}
=-\frac{1}{6}(\rho +3p).
\label{eqst}
\end{equation}
Assuming $\phi=\phi(t)$ for cosmological purpose, then the energy
density and pressure are given by, respectively,
\begin{equation}
\rho= \frac{1}{2} \dot \phi^2 +V(\phi),~~~~p= \frac{1}{2} \dot \phi^2
-V(\phi).
\label{Qep}
\end{equation}
The corresponding equation of state takes the form
\begin{equation}
\omega \equiv \frac{p}{\rho}= \frac{ \dot \phi^2 -2V(\phi)}{ \dot \phi^2 +2V(\phi)}.
\label{Qeqs}
\end{equation}
The causality restricts $\omega$ to be $|\omega|\le 1$. For $\dot \phi^2=0$, we
have a trivial case of $p=-\rho$, which implies that $\rho$ is
independent of $a$. Alternatively $V(\phi)$ plays the role of  a cosmological
constant.
On the other hand, if $V(\phi)=0$, one finds an extreme case of  a
massless scalar which plays the role of a stiff matter as $\rho $ drops as $1/a^6$.

The relevant equation of state  ranges over $-1<\omega<-1/3$, depending on the
dynamics of the field.
We note that when $ \dot \phi^2 < V(\phi)$ on later time,
we obtain  an accelerating universe of $p<-\rho/3$ from the second equation in Eq.(\ref{eqst}).
It was shown that for a flat spacetime $k=0$, $\phi$ and $\rho$ scale
as\footnote{Here our case of $\frac{\partial \phi}{\partial a}$ differs from
 $\frac{\partial \phi}{\partial a}= \frac{\sqrt {6(1+\omega)}}{a}$ in \cite{RP}
because our action Eq.(\ref{Qact}) is different from
\cite{RP}.}
\begin{equation}
\frac{\partial \phi}{\partial a}= \frac{\sqrt {3(1+\omega)}}{a},~~~
\rho \sim \frac{1}{a^{3(1+\omega)}}.
\end{equation}
Using the relation of $V(\phi)=(1-\omega) \rho/2$ with  $\phi=\sqrt{3(1+\omega)}\ln a$,
one of relevant cases\footnote{ There are three types of quintessence model :
(1) an inverse-power-law potential, (2) an exponential potential, and
(3) kinetic-term quintessence model\cite{MMA}.} which give the
quintessence is an exponential potential
\begin{equation}
V(\phi)= V_0 e^{-\sqrt{3(1+\omega)} \phi},~~~0< 3(1+\omega)<2
\label{Qpot}
\end{equation}
which is a kind of Liouville-type potential that decreases as $\phi$ increases.
According to the theory of quintessence, the dark energy of the
universe is dominated by the potential of a scalar field $\phi$
which is still rolling to its minimum at $V=0$. In addition we
require its minimum at $\phi=\infty$. The above Liouville-type
potential is suited well  for the quintessence. For example, if
one takes $\omega=-\frac{1}{2}<-\frac{1}{3}$, $V(\phi)=V_o e^{-\sqrt{3/2} \phi}$
derives an accelerating universe.
However, this model has a serious drawback which shows that it
leads to an eternally accelerating universe and accordingly to the
universe with an event horizon.

\section{ Quintessence within brane cosmology}
\label{sec-randall}

Now we wish to discuss issues of the quintessence within the brane cosmology.
For simplicity we choose the MDW approach.
For the cosmological embedding, a relevant solution is initially chosen as
the 5D topological anti de Sitter
(TAdS$_5$) black holes\cite{BIR} for right hand side $(+)$ and left hand side $(-)$,
\beq
ds^{2}_{5\pm}=g_{MN} dx^Mdx^N=-h_\pm(r)dt^2 +\fr{1}{h_\pm(r)}dr^2 +r^2
\gamma_{ij}dx^idx^j,
\label{BMT}
\eeq
where the metric function $h_\pm$ is given by
\beq
h_\pm(r)=k-\fr{\alpha_\pm}{r^2}+ \fr{ r^2}{\ell^2}
\eeq
with $\ell$ the curvature radius of AdS$_5$ spaces.
$\gamma_{ij}$ is the horizon metric of  a constant curvature manifold ${\bf M}^3$
with volume $Vol({\bf M}^3)= \int d^3x \sqrt{\gamma}$. The horizon
geometry of the TAdS$_5 $black hole is elliptic, flat, and hyperbolic for
$k=1,0,-1$, respectively\footnote{ On later, $k$ will turn out to be the
spatial curvature of the universe.}
:
$\gamma_{ij}dx^idx^j=\left[d\chi^2 +f_{k}(\chi)^2(d\theta^2+ \sin^2 \theta d\phi^2)
\right]$ with $ f_{1}(\chi) =\sin \chi,~f_{0}(\chi) =\chi, ~f_{-1}(\chi) =\sinh
\chi$.
In the case of $\alpha_\pm=0$, we have two exact AdS$_5$-spaces with the same
cosmological constants $\Lambda_{\pm}= -6/\ell^2$.  However,
$\alpha_\pm \not=0$ generates the electric part of the Weyl tensor $C^\pm_{MNPQ}$
on each side. Its presence  means that the bulk
spacetime has the two black holes with horizon located at $r=r_{\pm h}$,
where $r_{\pm h}^2=\ell^2(-k +\sqrt{k^2 +4
\alpha_\pm/\ell^2})/2$\cite{KRA}.
Now we introduce the location of brane (moving domain wall) in the form of
the radial, timelike geodesics parametrized by the proper time $\tau$
: $(t,r,\chi,\theta,\phi)\to(t(\tau),a(\tau), \chi, \theta, \phi)$.
Then the induced metric of dynamical domain wall will be given
by the conventional FRW-type.
In this case $\tau$ and $a(\tau)$ denote the cosmic time and scale factor of
the FRW universe, respectively. The tangent vectors of this brane
can be expressed as
\beq
u_\pm= \dot t_\pm \fr{\partial}{\partial t_\pm}+ \dot a \fr{\partial}{\partial
a},
\eeq
where the overdot means differentiation with respect to
$\tau$. These satisfy $u_{\pm M}u_{\pm N} g^{MN}=-1$.
Further we need the normal 1-forms
 directed toward to each side: $n_{\pm M}n_{\pm N} g^{MN}=1$. Here we choose these as
\beq
n_\pm=\pm \dot a dt_\pm \mp \dot t_\pm da.
\eeq
This embedding reduces to the Randall-Sundrum case when
$\alpha_\pm=0$\cite{RS2}. Using these, we can express $\dot t$ in terms of $\dot a$
as
\beq
\dot t=\fr{(\dot a^2 +h_\pm(a))^{1/2}}{h_\pm(a)}.
\label{TAV}
\eeq
From the bulk metric Eq.(\ref{BMT}) together with Eq.(\ref{TAV}), we arrive at the
4D induced metric
\beqa
ds^{2}_{4}&&=-d \tau^2 +a(\tau)^2
\left[d\chi^2 +f_{k}(\chi)^2(d\theta^2+ \sin^2 \theta d\phi^2)
\right]
\nonumber \\
&&\equiv h_{\mu \nu}dx^{\mu} dx^{\nu}.
\label{INM}
\eeqa
Hereafter we use the Greek indices ($\mu,\nu,\cdots)$  for  physics of the brane.
The extrinsic curvatures  for cosmological embedding are given by

\beqa
&&(K_\pm)_{\tau\tau}\equiv (K_\pm)_{MN} u^M_\pm u^N_\pm =
\pm(h_\pm \dot t_\pm)^{-1}(\ddot a +h'_\pm /2),
\label{Ktt}\\
&&(K_\pm)_{\chi}^{\chi} = (K_\pm)_{\theta}^{\theta}=(K_\pm)_{\phi}^{\phi}
=\mp h_\pm \dot t_\pm /a,
\label{Kss}
\eeqa
where the prime stands for  derivative with respect to $a$.
The above equation implies that the extrinsic curvature will jump  across the brane
if  a localized matter resides on the brane.
This jump is easily realized through the Israel junction condition.
 Here we need only its 4D induced version defined through Eqs.
 (\ref{Ktt}) and (\ref{Kss})
\beq
\triangle K_{\mu \nu}=-\kappa_5^2 \left(
T_{\mu\nu}-\fr{1}{3}T^{\lambda}_{\lambda}h_{\mu\nu} \right)
\label{4DI}
\eeq
with $\triangle K_{\mu \nu}\equiv (K_+)_{\mu \nu} -(K_-)_{\mu
\nu}$ and the 5D gravitational constant
$\kappa_5^2\equiv \kappa^2_4\ell=\ell$ when $\kappa_4^2=1$.
We usually  choose the localized stress-energy tensor on the
brane as the 4D perfect fluid

\beq
T_{\mu \nu}=(\varrho +p)u_{\mu}u_{\nu}+p\:h_{\mu\nu}
\label{MAT}
\eeq
with $\varrho=\rho_m+ \sigma(p=p_m-\sigma)$. $\rho_m(p_m)$ is the energy density (pressure)
of the localized matter and $\sigma$ is the brane tension
(or vacuum energy like cosmological constant).
In the absence of a localized matter,
Eq.(\ref{4DI}) takes the form of  $-\fr{\sigma \kappa_5^2}{3}
h_{\mu\nu}$ as in the RS case.
 We stress again that  $u_\mu$ are defined through
 the 4D induced metric $h_{\mu\nu}$ of Eq.(\ref{INM}). In addition, we need the
 Gauss-Codazzi equations with the 4D induced Einstein tensor $G_{\mu\nu}$
 with respect to $h_{\mu\nu}$

\beqa
&&\fr{\kappa_5^2}{2} \left[(K_+)_{\mu\nu}+(K_-)_{\mu\nu}
\right] T^{\mu\nu}=\triangle G_{\mu\nu} n^\mu n^\nu,
 \label{GCE1} \\
&&\kappa_5^2 {h_{\mu}}^{\lambda} \nabla_{\nu}{T_{\lambda}}^{\nu}={h_{\mu}}^{\lambda}
\triangle G_{\lambda}^{\nu} n_{\nu},
\label{GCE2}
\eeqa
where the last one is here nothing but the conservation law
\beq
{d \over d\tau} \left( \varrho a^3 \right)+p{d \over d\tau} \left( a^3
\right)=0.
\eeq
We note that  Eqs.(\ref{GCE1}) and (\ref{GCE2}) describe effectively the metric
junction condition. In
these equations,  right-hand sides become
zero because we choose the same cosmological constant for the two
sides.
From Eqs.(\ref{4DI}), (\ref{MAT}) and (\ref{GCE1}), one finds

\beqa
&&(h_{+} \dot t_{+})^{-1}(\ddot a +h'_{+} /2)+
(h_{-} \dot t_{-})^{-1}(\ddot a +h'_{-} /2)=
-\kappa_5^2 \left (p+\fr{2}{3} \varrho \right),\label{TPC} \\
&&h_{+}\dot t_{+}+h_{-}\dot t_{-}=\fr{\kappa_5^2}{3}\varrho a,\label{SPC} \\
&&(h_{+} \dot t_{+})^{-1}(\ddot a +h'_{R} /2)-
(h_{-} \dot t_{-})^{-1}(\ddot a +h'_{-} /2)=
3\fr{p}{\varrho a} \left (h_{+}\dot t_{+}- h_{-}\dot t_{-}
\right).
\eeqa
Equation (\ref{TPC}) corresponds to the acceleration  of the
moving domain wall, while Eq.(\ref{SPC}) corresponds to the
velocity of the moving domain wall.
We  solve the above equations simultaneously to give the second Friedmann
equation
\beq
\frac{ \ddot a}{a}=   -{1\over \ell^2} -\fr{\alpha_{+}+\alpha_{-}}{2a^4}
-\frac{\kappa_5^4}{18} \varrho( \varrho + \frac{3}{2}p) -{27\over4}
\fr{(\alpha_{+}-\alpha_{-})^2p}{\kappa_5^4 \varrho^3 a^8}.
\label{SFR}
\eeq
and the first Friedmann equation with the spatial curvature of the universe $k=1,0,-1$

\beq
\frac{ \dot a^2}{a^2}= -\frac{k}{a^2}  -{1\over \ell^2} +\fr{\alpha_{+}+\alpha_{-}}{2a^4}
+{1\over36} \kappa_5^4 \varrho^2  +{9\over4}
\fr{(\alpha_{+}-\alpha_{-})^2}{\kappa_5^4 \varrho^2 a^8}.
\label{FFR}
\eeq
For simplicity,
we choose hereafter the Z$_2$-symmetric case with $\alpha_+=\alpha_-\equiv
\alpha$. Furthermore we require the fine-tuning of $\sigma=6/\kappa_5^2\ell^2$
for two-sided brane world scenario. Then considering $\varrho=\rho_m+ \sigma(p=p_m-\sigma)$
with $\kappa_5^2=\ell$,
 Eq.(\ref{FFR})  leads to
\beq
\frac{ \dot a^2}{a^2}= -\frac{k}{a^2} + {\rho_m \over 3}
 +\fr{\alpha}{a^4}
+{\ell^2 \over 36} \rho_m^2.
\label{FFR1}
\eeq
The second Friedmann equation takes an interesting form with $p_m= \omega \rho_m$
\beq
\frac{ \ddot a}{a}=  -\fr{\rho_m}{6}
(1+3 \omega) -\frac{\alpha}{a^4}
-\frac{\ell^2}{36} \rho_m^2( 2 + 3\omega).
\label{SFR1}
\eeq

\subsection {$\alpha=0$ case}

In the case of $\alpha=0$, the bulk spacetime consists of  two
exact AdS$_5$ spacetimes. Let us first discuss its cosmological
implication by introducing a localized scalar with the potential $V(\phi)$.
The corresponding energy-stress tensor is also given by Eq.(\ref{Qst}).
In this case the energy density (pressure) of the localized matter
are given by the same form as in Eq.(\ref{Qst})
\begin{equation}
\rho_m= \frac{1}{2} \dot \phi^2 +V(\phi),~~~~p_m= \frac{1}{2} \dot \phi^2
-V(\phi).
\label{Lst}
\end{equation}
Expressing Eq.(\ref{SFR1}) in terms of Eq.(\ref{Lst}) leads to
\beq
\frac{ \ddot a}{a}=  -\fr{1}{3}(\dot \phi^2 -V(\phi))
-\frac{\ell^2}{4\cdot 36}(\dot \phi^2 +2V(\phi))(5\dot \phi^2 -2V(\phi)).
\label{SFR2}
\eeq
Here we classify three cases.

(i) $5 \dot \phi^2/2 \le V(\phi)$: dominance of the potential energy.

In this case we have the equation of state
$-1<\omega \le-\fr{2}{3}$. Two terms in right hand side give us  positive
accelerations. And thus we have more acceleration than the standard
quintessence. Here it is important to remind the reader  that the observational evidence for a
cosmological constant is really the same bound of $-1<\omega_{observed}
\le-\fr{2}{3}$ as is here\cite{HKS}.
A borderline of $\omega=-1$ can be achieved when $\dot
\phi^2=0$. In this case the potential term $V(\phi)$ plays the role
of a cosmological constant.
 Hence we expect presumably that the quintessence within the
brane cosmology has something to tell us how to resolve the cosmological constant
problem.

(ii)  $\dot \phi^2 <V(\phi)< 5\dot \phi^2/2$.

Here we have
$-\fr{2}{3}<\omega<-\fr{1}{3}$. This is a very interesting case
because the first term induces  positive acceleration, while the
second induces negative acceleration. So there exists
competition between two terms and accordingly it may have a chance to
exit from the accelerating universe in the early universe.

(iii) $ \dot \phi^2 > V(\phi)$: dominance of the kinetic energy.

 The corresponding equation of state is
$-\fr{1}{3}< \omega< 1$. This corresponds to the deceleration
epoch because both two terms induce  negative accelerations.
The case of $\dot \phi^2=V(\phi)$ corresponds to a borderline of
$\omega=-1/3$.

We consider a case that
 the quadratic term $(\rho_m^2)$ in right hand
side of Eq. (\ref{FFR1})   becomes significant in comparison with the
linear term $(\rho_m)$ in the early universe. This is possible when $\rho_m>12/\ell^2$.
Considering $k=\alpha=0$ and $\rho_m^2$-term only
with $\rho_m \sim 1/a^{3(1+\omega)}$, Eq.(\ref{FFR1}) is solved to
give $a(\tau)\sim \tau^q$ with $q=\fr{1}{3(1+\omega)}>\fr{1}{2}$. On the other hand,
the conventional quintessence  with $\rho_m$
leads to  $a(\tau)\sim \tau^p$ with $p=\fr{2}{3(1+\omega)}>1$.
Hence when the quadratic term is dominated in the early universe, the Hubble expansion rate
decreases. In this  case the friction term (the second one) in
Eq.(\ref{seqa}) becomes small and the dynamics of the scalar field
will  be drastically changed from the conventional quintessence even for the
same potential\cite{MMA}. It is easily expected that a kinetic term will
play a more important role than the potential. Assuming this
scenario,  we could have a chance to meet the case (iii): dominance of the kinetic
term. Then it is proposed that the universe may show a nice
transition from an accelerating universe to a decelerating one
 in the early universe\cite{FKMP}.

\subsection{$\alpha \not=0$ case}

In this case we have  the holographic matter term from the bulk
configuration\footnote{In the BDL approach, we can also obtain this term\cite{KLMK}.}.
 This term of $\alpha/a^4$ in Eqs.(\ref{FFR1})
and (\ref{SFR1})
originates from the masses of the topological  black holes.
It always behaves like radiation  for  either a dark matter
candidate\cite{KRA,MMA} or
a CFT-matter\cite{VER}.
Here we are interested especially in a flat universe ($k=0$) with a CFT-radiation
matter\cite{Youm}.  We have  $\alpha=\omega_4 M$,
where $\omega_4= \fr{2\kappa_5^2}{3 Vol({\bf M}^3)},
M=\fr{a}{\ell}E, \kappa_5^2=\ell$. Here $M$ is the ADM mass of the TAdS$_5$
 black hole and $E$ is its holographic energy on the brane.
 Further $ V=a^3 Vol({\bf M}^3)$ is the size of the universe.
  Then one gets a universe filled partly  by the  CFT-radiation
 and  the scalar matter. Its acceleration is given by
\beq
\frac{ \ddot a}{a}=   -\fr{\rho_{CFT}}{6} -\fr{\rho_m}{6}(1
+3 \omega)
-\frac{\ell^2}{36} \rho_m^2( 2+ 3\omega),
\label{SFR3}
\eeq
where $\rho_{CFT}=E/V$ is the energy density of the CFT-holographic
matter.
Another form of the
above equation is
\beq
\frac{ \ddot a}{a}=  -\fr{1}{3}(\fr{1}{2}\rho_{CFT}+ \dot \phi^2 -V(\phi))
-\frac{\ell^2}{4\cdot 36}(\dot \phi^2 +2V(\phi))(5\dot \phi^2 -V(\phi)).
\label{SFR4}
\eeq
Here it is important to note that  energy densities in Eq.(\ref{SFR3}) drop like
\beq
\rho_{CFT}= \fr{\alpha}{a^4},~~~
\rho_m \sim \fr{1}{a^{3(1+\omega)}},~~~
\rho_m^2 \sim \fr{1}{a^{6(1+\omega)}},~~~0< 3(1+\omega)<2.
\eeq
Considering  the CFT-term $(a_{CFT}(\tau)\sim \tau^{1/2})$,
this term dominates in the very early universe\cite{FKMP}.
This is because $ a_{\rho_m} \sim \tau^{p}$ with $p>1$ and
$a_{\rho^2_m} \sim \tau^{q}$ with $q>1/2$. On later the quadratic
term will be significant because we expect
a transition from radiation to matter: $\fr{1}{2}\to q \to p$ as the universe
evolves.  If we have a phase of dominant kinetic term during this transition, we can
apply the same scenario as in the $\alpha=0$ to this case to have
a nice exist from the  accelerating universe.

\section{Discussions}

We discuss  issues of the  quintessence in the brane cosmology.
The initial bulk spacetime consists of the  two 5D topological anti de Sitter
black hole joined by the brane. Considering the embedding of a moving domain wall
into this background, we find the FRW universe with a scale factor
$a(\tau)$. Here this factor is determined by not the Einstein equation in
Eq.(\ref{Qeq}) but the Israel junction condition Eq.(\ref{4DI}) mainly.
For our purpose we do not introduce any conventional radiation and matter.
Instead we include a localized scalar with a potential  on the brane as a stress-energy tensor,
and thus  we find  the quintessence  which gives an accelerating
universe. Importantly, we obtain  both
a quadratic term of $\rho^2$, as a representative of the brane cosmology
and a holographic matter term of $\alpha/a^4$ from the
topological black holes. We suggest a possible scenario that in the early universe,
$\rho^2$-term makes a large kinetic term which induces a
decelerating universe. This may provide a way to exit  from the
accelerating universe in the early universe.
Furthermore, one finds that if a holographic CFT-matter from the TAdS$_5$ black holes includes,
it will plays the role of a radiation  in the very early universe.

In this work we do not choose any explicit form of the potential
$V(\phi)$ which is suitable for explaining the early accelerating universe.
 There exist three kinds of potential for the present-day accelerating universe: (i)
Liouville-type
potential : $V(\phi)\sim \exp(-\sqrt{3(1+\omega)}\phi)$.
(ii) inverse power law potential : $V(\phi) \sim 1/\phi^n$.
(iii) exponential potential with $1/\phi$ : $V(\phi)\sim (\exp
(M_P/\phi)-1)$. According to authors in\cite{FKMP}, solutions of all these
models asymptote towards the equation of state $p=-\rho$, showing
eternally accelerating universe. This type of the universe (like de Sitter spacetime) has the
future horizon and thus is not suited to an S-matrix or S-vector
description\cite{HKS}.

We compare our model with Mizuno and Maeda (MM) case\cite{MMA}.
As we were almost finishing our work, we are informed that their
paper discussed similar issue in the brane cosmology.
Both two models are suitable for explaining a transition from an
accelerating universe to the standard one. But there exist differences between
MM and our cases. The first one is that we do not include any
conventional radiation and matter. Also we introduce an explicit
form of a holographic matter from the TAdS$_5$ black holes.
This is regarded as a CFT-radiation matter\cite{VER}. But they
neglected this in favor of conventional radiation and matter.

Finally we summarize our result.
Initially this work aims for resolving an eventually accelerating universe of
the present-day quintessence.
However, the brane cosmology provides a
mechanism to exit from an accelerating universe only in the early
universe. In this case the relevant term is $\rho_m^2$-term which
makes a large kinetic term. This induces a transition from an
accelerating universe to a decelerating universe in the early universe.
Although we fail to resolve the problem of the present-day
quintessence, we expect that this approach provide a hint of
avoiding an eternally accelerating universe, assuming our universe is presently
accelerating.

\section*{Acknowledgement}
We thank to C. R. Cai and H.W. Lee for helpful discussions.
This work was supported in part by the Brain Korea 21
Program of  Ministry of Education, Project No. D-1123 and
 KOSEF, Project No. 2000-1-11200-001-3.

\end{document}